\documentclass[aps,prb,twocolumn,showpacs,superscriptaddress]{revtex4-1}

\pdfoutput=1

\usepackage{graphicx}
\usepackage{dcolumn}
\usepackage{bm}
\usepackage{color}
\usepackage{amsmath}
\usepackage{float}
\usepackage{verbatim}   					 
\usepackage{subfigure}  				 
\usepackage{enumerate}  				 
\usepackage{placeins}   					 
\usepackage{booktabs}
\usepackage{epstopdf}
\usepackage{chngcntr}
\usepackage{array}
\usepackage{graphicx}

\newcommand{\subfigimg}[3][,]{%
  \setbox1=\hbox{\includegraphics[#1]{#3}}
  \leavevmode\rlap{\usebox1}
  \rlap{\hspace*{0pt}\raisebox{\dimexpr\ht1-1\baselineskip}{#2}}
  \phantom{\usebox1}
}

\begin{document}

\newcommand{\bra}[1]{\left\langle#1\right|}
\newcommand{\ket}[1]{\left|#1\right\rangle}
\newcommand{\bracket}[2]{\big\langle#1 \bigm| #2\big\rangle}
\newcommand{\PPV}{\ket{0_{\rm PP}}}
\newcommand{\Tr}{{\rm Tr}}
\renewcommand{\Im}{{\rm Im}}
\renewcommand{\Re}{{\rm Re}}
\newcommand{\MC}[1]{\mathcal{#1}}
\newcommand{\p}{{\prime}}
\newcommand{\pp}{{\prime\prime}}
\newcommand{\ppp}{{\prime\prime\prime}}
\newcommand{\pppp}{{\prime\prime\prime\prime}}
\newcommand{\TK}{T_{\rm K}}

\title{Electronic Transport in Gadolinium Atomic-Size Contacts}

\author{B. Olivera}
\affiliation{Departamento de F\'isica Aplicada and Unidad Asociada del CSIC, Facultad de Ciencias, Universidad de Alicante, San Vicente del Raspeig, E-03690, Spain.}
\author{C. Salgado}
\affiliation{Departamento de F\'isica de la Materia Condensada, Universidad Aut\'onoma de Madrid, E-28049 Madrid, Spain.}
\author{J. L. Lado}
\affiliation{International Iberian Nanotechnology Laboratory (INL), 4715-330 Braga, Portugal.}
\author{A. Karimi}
\affiliation{Department of Physics, University of Konstanz, Universit\"{a}tsstra{\ss}e 10, D-78464 Konstanz, Germany.}
\author{V. Henkel}
\affiliation{Department of Physics, University of Konstanz, Universit\"{a}tsstra{\ss}e 10, D-78464 Konstanz, Germany.}
\author{E. Scheer}
\affiliation{Department of Physics, University of Konstanz, Universit\"{a}tsstra{\ss}e 10, D-78464 Konstanz, Germany.}
\author{J. Fern\'andez-Rossier}
\affiliation{Departamento de F\'isica Aplicada and Unidad Asociada del CSIC, Facultad de Ciencias, Universidad de Alicante, San Vicente del Raspeig, E-03690, Spain.}
\affiliation{International Iberian Nanotechnology Laboratory (INL), 4715-330 Braga, Portugal.}
\author{J. J. Palacios}
\affiliation{Departamento de F\'isica de la Materia Condensada, Condensed Matter Physics Center (IFIMAC), and
Instituto Nicol\'as Cabrera, Universidad Aut\'onoma de Madrid, E-28049 Madrid, Spain.}
\author{C. Untiedt}
\affiliation{Departamento de F\'isica Aplicada and Unidad Asociada del CSIC, Facultad de Ciencias, Universidad de Alicante, San Vicente del Raspeig, E-03690, Spain.}

\date{\today} 

\begin{abstract} 
We report on the fabrication, transport measurements, and density functional theory (DFT) calculations of atomic size contacts made out of gadolinium (Gd). Gd is known to have local moments mainly associated with $f$ electrons. These coexist with itinerant $s$ and $d$ bands that account for its metallic character. Here we explore whether and how the local moments influence electronic transport properties at the atomic scale. Using both Scanning Tunneling Microscope (STM) and lithographic Mechanically Controllable Break Junction (MCBJ) techniques under cryogenic conditions, we study the conductance of Gd when only few atoms form the junction between bulk electrodes made out of the very same material. Thousands of measurements shows that Gd has an average lowest conductance, attributed to an atom-size contact, below $\frac{2e^2}{h}$.  Our DFT calculations for monostrand chains anticipate that the $f$ bands are fully spin polarized and insulating, and that the conduction may be dominated by $s$, $p$, and $d$ bands. DFT quantum transport calculations quantitatively reproduce the experimental results for zero bias and reveal that,  while $s-p$ bands are dominant for transport, $d$ orbitals seem to have a relevant contribution in some cases.
\end{abstract}


%
%

\maketitle

\section{Introduction}

Quantum transport plays a key role in the electrical response of  
atomic scale contacts, giving rise to new phenomena differing from the bulk behaviour of the different materials\cite{Yanson01QuantumMechanismsNanowires,Agrait03QuantumPropertiesAtomicSizedConductors,Naidyuk05PointContactSpectroscopy}. The central assumptions that permit to have a first guess of the conductance of an atomic scale contact are two. First, the conductance of the system is determined by the elastic transmission of the electrons at the Fermi level (Landauer formalism) and, second, the number of transmission channels that appear in the Landauer formula is determined by the chemical valence of the atoms\cite{Scheer98ChemicalValenceConductionSingleAtomContact}. 

After three decades of exploration of electronic transport in atomic scale contacts, many materials with different physical properties have been studied. 
The groups that are relatively well understood include noble metals, such as Au\cite{Agrait93ConductanceStepsQuantizationAtomicSizeContacts} and Pt\cite{Smit01CommonOriginReconstructionChains}, $sp$ metals, such as Al\cite{Scheer97ConducChannTransAtomSizeAlContacts,SanchezPortal97NanocontactsElecStructExtremeUniaxStrainAluminum} and Zn\cite{Hafner2004ConducChannOneAtomZnContacts}, ferromagnetic $3d$ transition metals, such as Fe, Co and Ni\cite{Calvo09KondoEffectFerromagneticAtomicContacts}, superconductors such as Pb\cite{Yanson01QuantumMechanismsNanowires,Muller2016PlasticitySingleAtomPbJunctions,Cuevas1998EvolConducChanMetalAtContacElastDeform}, and even semi-metals such as Bi\cite{Sabater13TopologicallyProtectQuantTransLocalExfolBiRoomTemper,Pernau2014MagnetotranAtomSizeBiContacts}. Besides, there are some metals like Ir\cite{Smit01CommonOriginReconstructionChains}, Pt\cite{Smit01CommonOriginReconstructionChains} and Au\cite{Yanson1998FormManipulMetalWireSingleAuAtoms,Ohnishi1998QuantCondIndividRowsSuspenAuAtoms} that form long chains of atoms. Still, despite of all this effort, some important families remain to be covered.

In this context, while atomic contacts  with $s$, $p$,  and $d$ electrons  have been widely explored,  systems with partially filled $f$ shells remain pretty much an uncharted territory (with a few  exceptions\cite{Muller10SwitchingConducDyNanocontactsMagnetostriction,
Jammalamadaka15RemotContrlMagnetostricNanocontRoomTemper}). 
On another hand, there has been an interest to unveil the role of magnetism in the electronic transport in atomic-sized contacts. Later attempts in $d$ materials have shown Kondo screening of the magnetic moments at such scale\cite{Madhavan98SpectroscopicEvidenceKondoResonance,Calvo09KondoEffectFerromagneticAtomicContacts,Calvo12AnalysisKondoEffectFerromagneticAtomicSizedContacts}. $f$ materials are therefore also  good candidates to study the influence of the $f$ decoupled magnetic moments on the transport electrons, mainly of $s-p$, and maybe $d$ character. 

Gd is a rare earth metal that belongs to the lanthanide group with the electronic configuration [Xe] $4f^7 5d^1 6s^2$. It is a trivalent metal\cite{Gschneidner78HandbookRareEarth} that in bulk is a strong ferromagnet with $T_C=293.2\,\mathrm{K}$\cite{Nigh63MagnetElecResistGdSingleCrystals} with hexagonal close-packed (hcp) structure. It presents interesting properties, such as very high neutron absorption\cite{Lapp1947NeutronAbsorbIsotGdSm,Leinweber2006NeutronCaptureTotalCrossSectMeasResonParamGd} and a pronounced magnetocaloric effect \cite{GschneidnerJr2005RecentDevelopMagnetocalMater}. Regarding other type of experimental measurements on rare earths, studies of electron-magnon interaction on point contacts made out of Gd, holmium (Ho), and terbium (Tb)\cite{Akimenko81PCSMagnonsMetals} as well as electronic structures measurements with photo-electron spectroscopy\cite{Bovensiepen07CoherentIncoherentExcitationsGd0001SurfUltrafast} have been performed.
There are few experimental works about electronic transport on rare earth atomic-size contacts. Some of them\cite{Berg14ElektrTransNanokontakteSeltenErdMetallen,Muller10SwitchingConducDyNanocontactsMagnetostriction,Jammalamadaka2015RemoteControlMagnetostricBasedNanocontRoomTemp} reported measurements on nanocontacts made out of metals such as yttrium (Y), cerium (Ce), dysprosium (Dy), and Gd itself by using notched-wire MCBJ technique\cite{Agrait03QuantumPropertiesAtomicSizedConductors}.

Concerning calculations on lanthanide materials, not much has been published for atomic-scale contacts but their bulk properties have been widely studied. Calculations of the magnetic moment\cite{Ahuja94ElecStructMagFermiSurfGdTb,Kurz02MagnetElecStrucHcpGdGd0001Surf} of bulk Gd compare well with the measured $7.63\,\mu _B$\cite{Nigh63MagnetElecResistGdSingleCrystals}, where approximately $7\,\mu _B$ come from the $4f^7$ orbital. As a result, the remaining $0.63\,\mu _B$ belong to the conduction electrons. Exchange interaction studies on Gd can be found elsewhere as well\cite{Goodings62MagnitudeExchangeInteractGd}. Moreover, several groups have calculated the electronic band structure\cite{Temmerman90BandModelGroundStateGd} from where the electronic density of states (DOS) as well as its projection on different orbitals has been inferred\cite{Turek03AbInitioExchangeInteracCurieTemperBulkGd,Santos04FerromagTemperDepElecStrucHcpGd,Duan07ElecMagneticTransportPropsRareEarthMonopnictides}.

Here we present a combined experimental-theoretical work with two independent experimental techniques along with DFT calculations. With STM we obtain a higher amount of statistical data than what can be obtained with lithographed MCBJ, which offers samples with much higher stability. DFT calculations of both electronic structure and transport properties have been carried out to shed light on the experimental results.

\section{Experiments}
\begin{figure}[!htb]
\begin{center}$
\begin{array}{c}
\includegraphics[width=3.in]{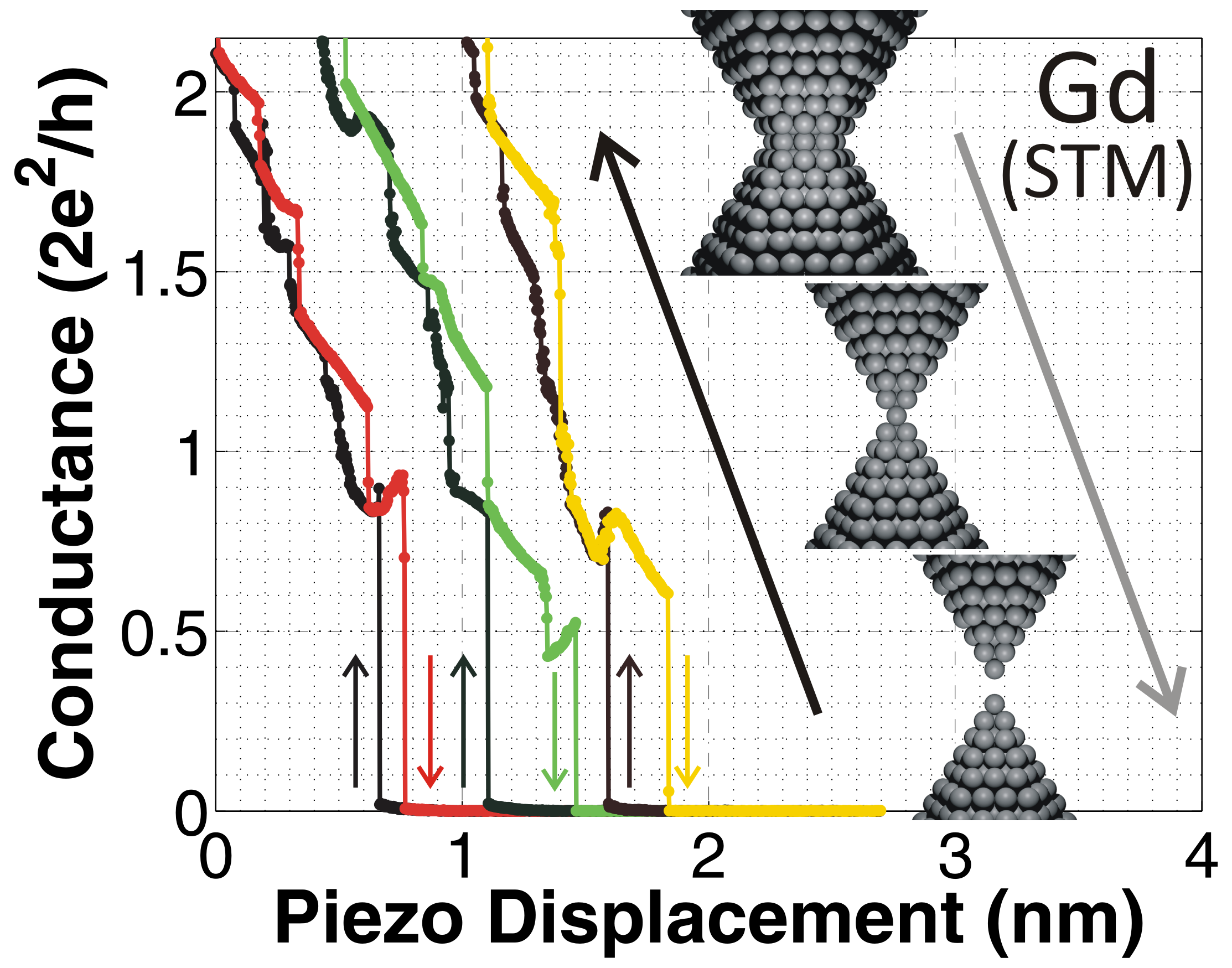} \\
\includegraphics[width=3.in]{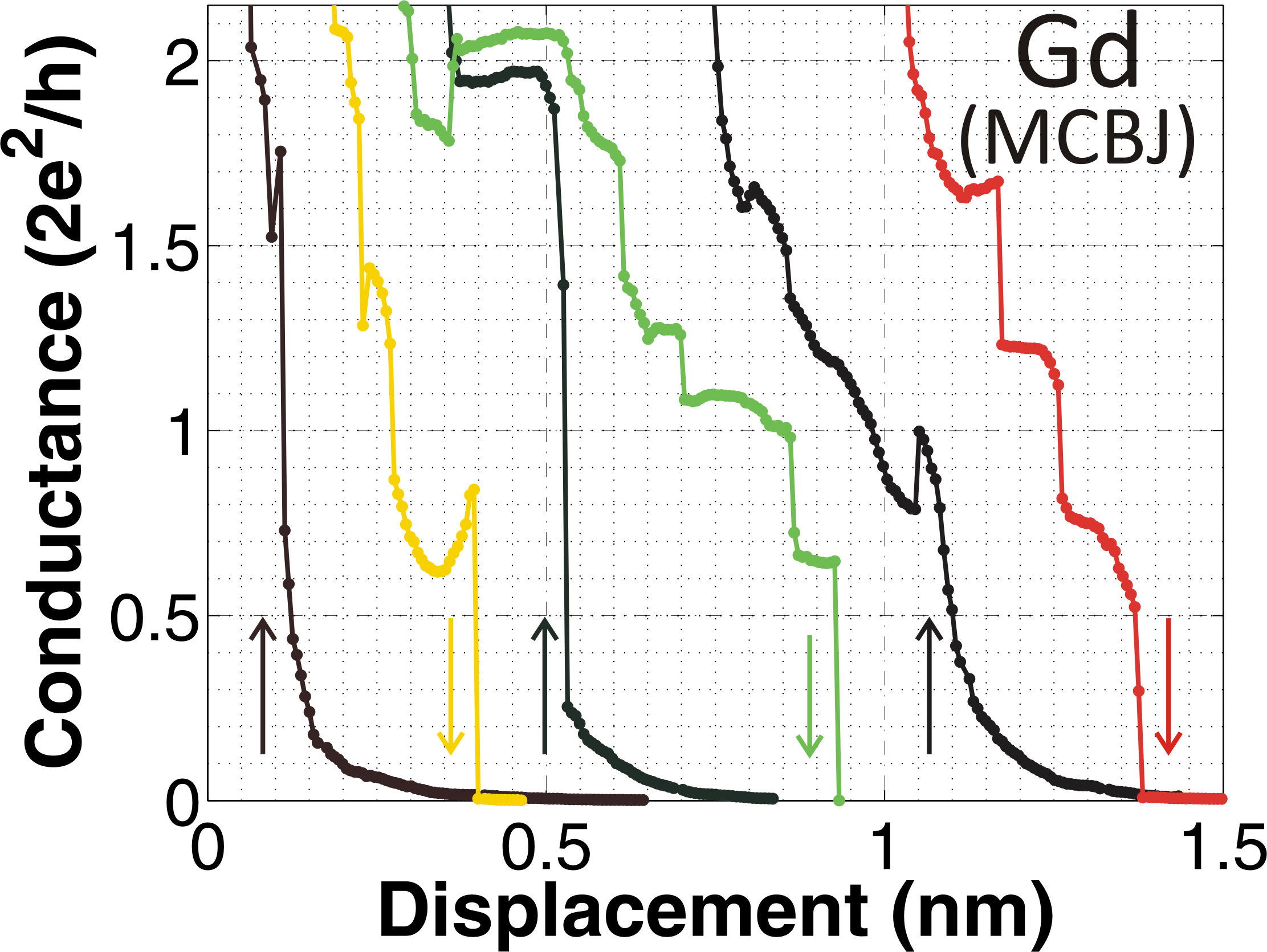}
\end{array}$
\end{center}
\caption{Typical conductance traces for atomic size contacts made out of Gd. Bright/dark curves stand for breaking/creating contacts. Upper plot: Measurements taken with STM technique in equilibrium with liquid He bath and at $10^{-8}\,\mathrm{mbar}$. All traces have been taken at $100\,\mathrm{mV}$ bias voltage. Inset: artistic representation of nanocontacts; hcp ball-stacking is pictured, where balls represent atoms. Lower plot: Measurements taken with lithographic MCBJ technique in equilibrium with liquid He bath and at $10^{-5}\,\mathrm{mbar}$. Red and green traces and their return traces have been taken at $5\,\mathrm{mV}$ bias voltage, yellow curve has been taken at $10\,\mathrm{mV}$.}
\label{fig:3_traces_Gd}
\end{figure}

Atomic-size contacts are the narrowest experimentally accessible junctions between bulk electrodes made out of the same material\cite{Gimzewski87TransitionTunnelPointContactSTM} (see inset in Fig. \ref{fig:3_traces_Gd}). In this work we build up nanocontacts made out of pure metallic Gd. In order to study electronic transport on nanocontacts, we use STM\cite{BinningRohrer82STM} and lithographic MCBJ\cite{Moreland85ElecTunnelingExperNbSnBreakJunctions,Ruitenbeek96NanofabricatedAtomicSizeContacts} techniques, independently. With both techniques we record the electrical current through nanocontacts under fixed applied DC bias voltage when changing the contact geometry.

We use STM technique in contact mode and we read the current from a low-noise amplifier with a gain factor of $5$. With piezoelectric materials we control the distance between bulk electrodes with atomic precision ($\sim 1\,\mathrm{pm}$) under cryogenic conditions (liquid helium bath). Samples consist of two wires of $\sim 1\,\mathrm{mm}$ diameter made out of the same material and cross-shaped arranged in order to avoid multi-contact locations. With this technique we build atomic contacts in a straightforward way, that is bringing into and out of contact the bulk wire-shaped electrodes by applying electrical DC sawtooth pulses to the piezoelectric materials mentioned above.

Besides, and for comparison purposes, we use the MCBJ technique\cite{Ruitenbeek96NanofabricatedAtomicSizeContacts} in which a motor moves a pushing rod with micro-metric precision. This rod bends the lithographed sample from the rear side of the substrate right below the nano-junction location. The fabrication process of the latter will be explained below. The movement of the rod is reversible, so that atomic contacts can be created and broken repeatedly and the electronic transport through them is measured in the same way as with the STM technique.

Gd gets quickly oxidized in contact with air. In order to avoid contact with environmental compounds and to preserve the purity of these materials we use the following methods. For STM experiments, we mount samples inside a custom-made controlled atmosphere chamber. Gd wire-shaped samples has $99.9\%$ purity and $0.5\,\mathrm{mm}$ diameter. We use argon gas ($99.999\%$ pure) as surrounding atmosphere before closing the STM under high vacuum conditions ($10^{-8}\,\mathrm{mbar}$ reached with turbo-molecular pumping). Besides, right before starting pumping, a ceramic (i.e. insulator and non-magnetic) knife is used for scratching the outer layer of Gd wires that are afterwards brought into contact. After pumping at room temperature, the STM is inserted into a bath cryostat filled with liquid helium (He). Then, when samples reach equilibrium with liquid He temperature and in order to ensure clean electrodes, we measure the conductance as a function of distance when approaching or retracting the electrodes, so called conductance (creating/breaking) traces. Thousands of conductance traces with deep indentations (beyond $100\,\frac{2e^2}{h}$) are recorded along with electro-migrative fast ($\approx0.5\,\mathrm{s}$) DC pulses of $10\,\mathrm{V}$ that are applied to randomly chosen atomic-size contacts.

In MCBJ experiments, we ensure the purity of Gd samples from the technique principle itself. Lithographic MCBJ measurements start with junctions of $\approx2\,\mu \mathrm{m} $ long, $\approx75\,\mathrm{nm}$ thickness, and $\approx100\,\mathrm{nm}$ width. For sample preparation, Gd pellets with $99.9\%$ purity are thermally evaporated from a tungsten (W) boat onto a lithographed substrate. We have chosen W as boat material in order to prevent alloy formation with Gd. The substrates consist of a stacked series of materials\cite{Ruitenbeek96NanofabricatedAtomicSizeContacts}. From bottom to top: bronze $\approx250\, \mu \mathrm{m}$ thick, polyimide (Durimid 115A) $\approx1\, \mu \mathrm{m}$ thick, MMA-MAA (methyl-methacrylate-co-methacrylacid) $\approx540\,\mathrm{nm}$ thick and PMMA (poly(methyl-methacrylate)) $\approx140\,\mathrm{nm}$ thick. After proper lithography (with Scanning Electron Microscope (SEM) technique), Gd is finally evaporated on top of the described substrate. In order to avoid possible oxide coming from pristine Gd pellets, we cover the substrate for the first couple of evaporated nanometres of material. After evaporation, a subsequent final etching with oxygen plasma (Reactive Ion Etching (RIE)) is performed to suspend nano-bridge. The sample is then mounted to the three-point bending mechanism anchored to a cryostat insert, pumped to a moderate high vacuum of $10^{-5}\,\mathrm{mbar}$, and cooled down to liquid He temperature. The MCBJ contact is broken for the first time, when $4.2\,\mathrm{K}$ are reached. In this way, the few nanometres thick outer layer of Gd oxide at the sample protects the pure Gd nano-junction before MCBJ measurements start.

\begin{figure}[!htb]
\begin{center}$
\begin{array}{c}
\includegraphics[width=3.in]{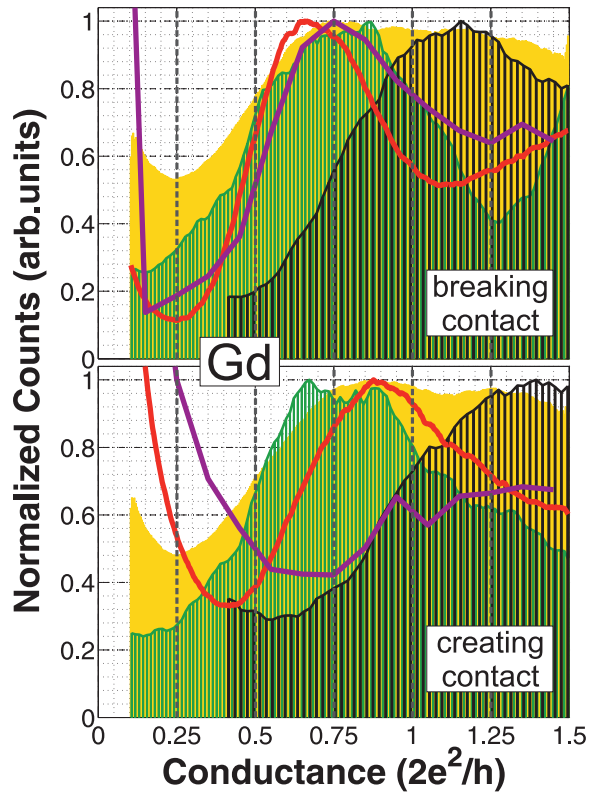} 
\end{array}$
\end{center}
\caption{Histograms of conductance out of traces for Gd atomic size contacts. Upper/lower plot stands for breaking/closing contact mode. Same colours for upper and lower plot stand for a set of traces with electrodes that have not been modified with deep indentations (i.e. less than $\approx20\,\frac{2e^2}{h}$). All measurements have been taken with STM technique at $100\,\mathrm{mV}$ bias voltage in equilibrium with liquid He bath and $10^{-8}\,\mathrm{mbar}$, except the purple curve, that has been taken with MCBJ technique, where a bias voltage of $10\,\mathrm{mV}$ has been applied. For STM histograms few (from $1$ to $10$) thousands of traces are considered. For MCBJ $\approx500$ traces are included.}
\label{fig:combined_hist_Gd_open_close_160805_01}
\end{figure}

Experimentally, we record the electrical conductance as a function of distance between bulk electrodes\cite{Agrait93ConductanceStepsQuantizationAtomicSizeContacts}, obtaining the so-called conductance traces (see Fig. \ref{fig:3_traces_Gd}). We focus on the last stages before breaking the contacts into the vacuum tunnel regime and the first ones when establishing metallic atomic size contacts. The resulting conductance is in the order of  $\frac{2e^2}{h}$,  as expected for a quantum conductor with a few channels, and shows abrupt changes as a function of the electrode distance that reflect variations in the atomic configuration of the nanocontact (see Fig. \ref{fig:3_traces_Gd}). With STM technique we manage to create stable low-noise traces of conductance at a rate of about $10$ traces per second. 
This allows us to obtain significant statistical data in a relatively brief period of time. The MCBJ technique enables mechanically more stable contacts than the STM one, but its rate of recording conductance traces is limited to about one trace per minute. We make histograms of conductance\cite{Yanson97DoHistProofCondQuantiz} out of the measured traces with STM and MCBJ (see Fig. \ref{fig:combined_hist_Gd_open_close_160805_01}). We concentrate here on few atom contacts, therefore only a window of few $\frac{2e^2}{h}$ (starting from zero) is considered.

We have recorded conductance traces for different electrode configurations. Every configuration comes from geometrical reconstruction of bulk electrodes by strong indentation of electrodes with STM technique (beyond $100\,\frac{2e^2}{h}$). With MCBJ, a smaller amount of data is obtained, therefore less variety of traces of conductance is achieved. Different families of conductance traces for this last technique are obtained moving back the pushing rod until reaching conductances beyond $50\,\frac{2e^2}{h}$.

\begin{figure*}[!htb]
\begin{center}$
\begin{array}{cc}
\includegraphics[height=2.3in]{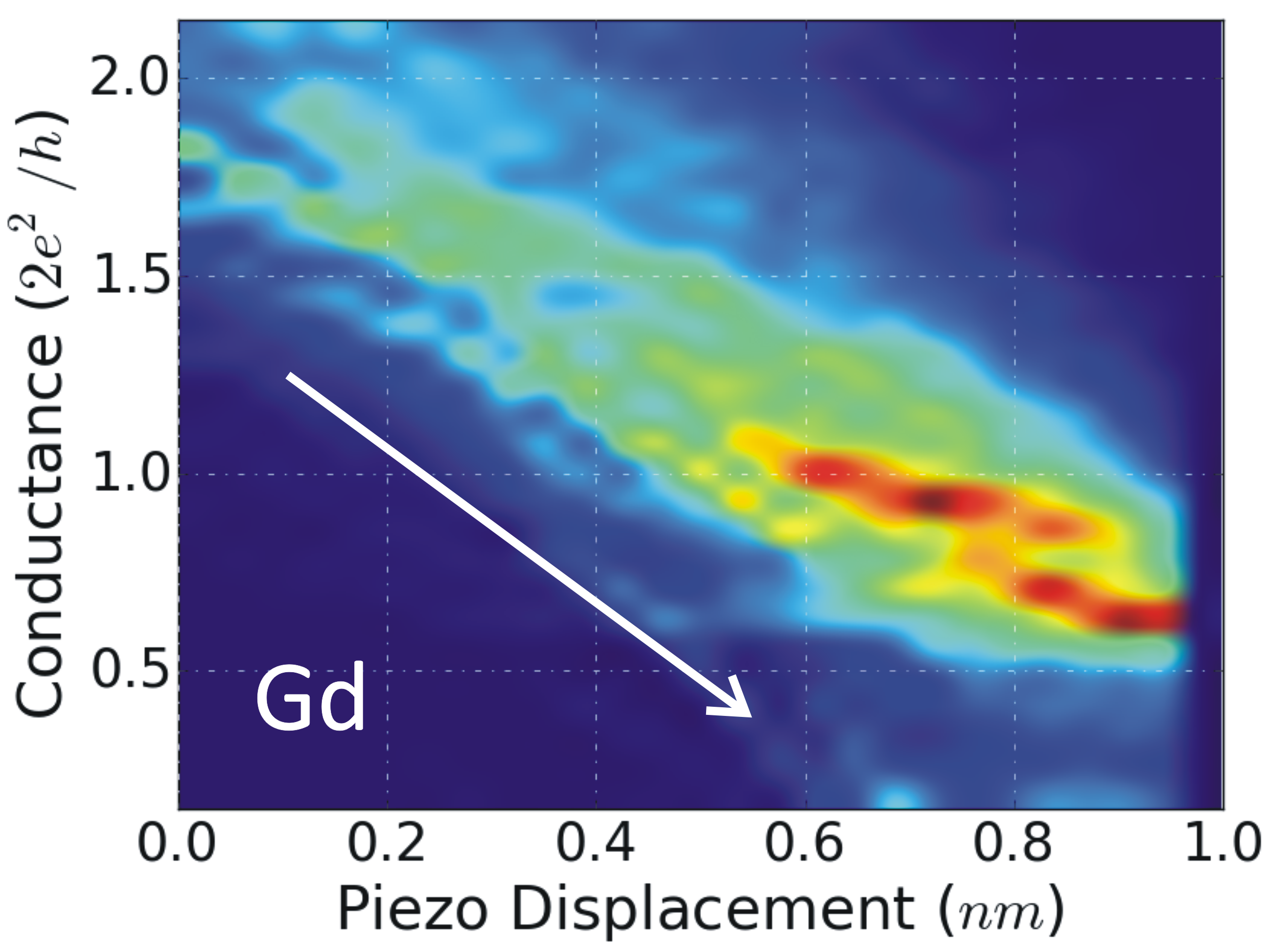} &
\includegraphics[height=2.3in]{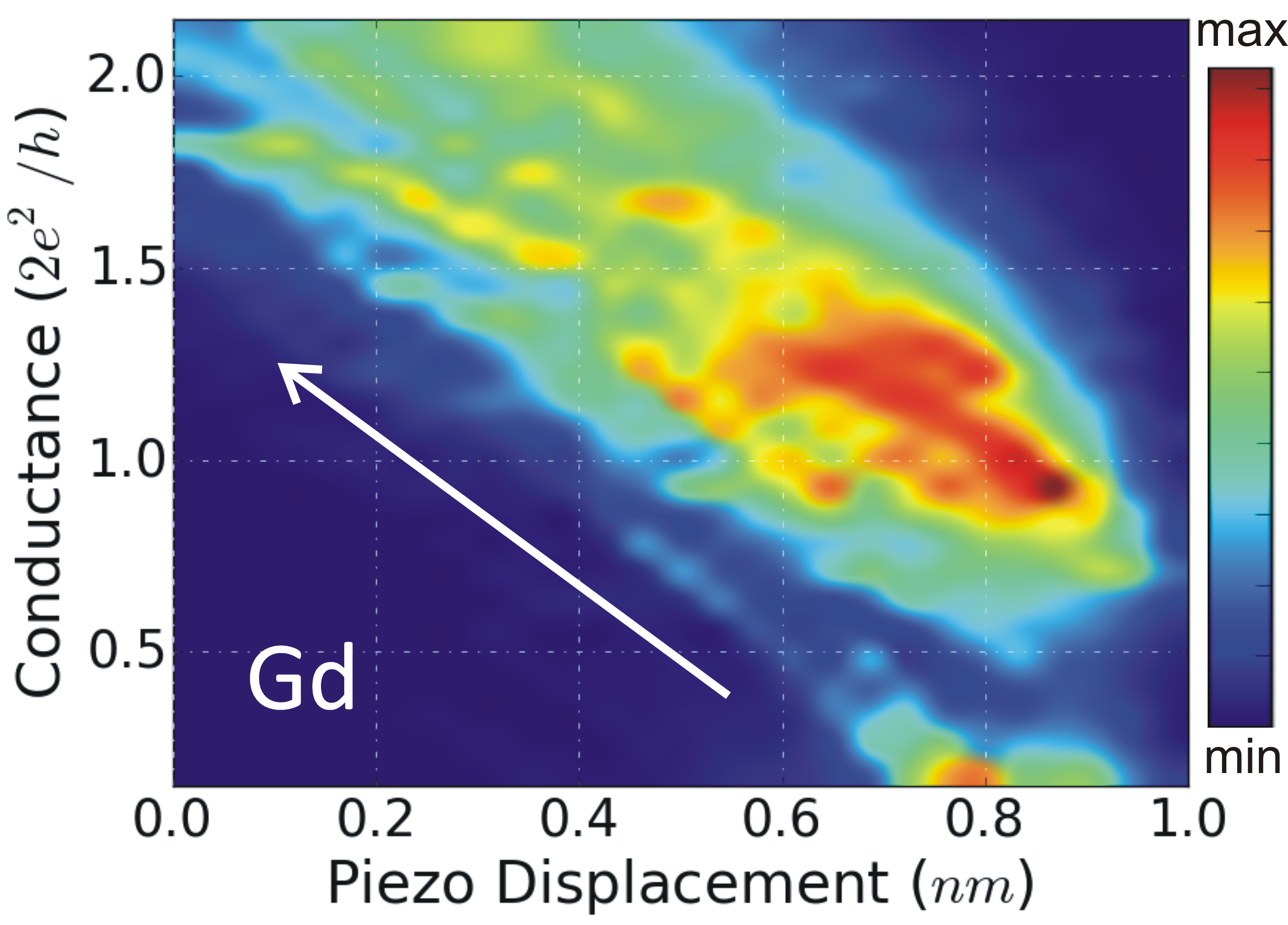} \\
\end{array}$
\end{center}
\caption{(Linear color scale) Overlapped Gd conductance traces measured with STM. Breaking contact (left-sided plot) and creating contact (right-sided plot) situations are shown. Traces are centred to $1\,\mathrm{nm}$ when $0.01\,\frac{2e^2}{h}$ is reached from higher/lower to lower/higher conductance values for breaking/creating contacts. Total number of traces are $2524$ and $2511$ for breaking and creating contact cases, respectively.}
\label{fig:overlapped_traces_Gd}
\end{figure*}

In Fig. \ref{fig:3_traces_Gd} we show typical conductance traces. In the STM case (upper plot), we show a series of measurements few seconds spaced between them where deep indentations are performed. We observe that the last plateau always falls at conductance values visibly smaller than $\frac{2e^2}{h}$. The same observation applies for MCBJ measurements (lower plot). As we will discuss below, this may be unexpected, because of the presence of a $s$ band at the Fermi energy, that normally contributes with one open channel per spin.  We see that the plateau shapes are negatively inclined, i.e. revealing lower conductance upon further stretching, as previously observed for other materials such as Pb\cite{Cuevas1998EvolConducChanMetalAtContacElastDeform}, but different from the observations of other soft metals, like Au. However, some of the last  plateaus reveal rising conductance upon stretching, as systematically observed for example in the case of Al\cite{Scheer97ConducChannTransAtomSizeAlContacts,SanchezPortal97NanocontactsElecStructExtremeUniaxStrainAluminum,Cuevas1998EvolConducChanMetalAtContacElastDeform}. Both effects, the falling and the rising last plateaus, are observed with both measurement techniques and are therefore attributed to intrinsic properties of Gd contacts. Another observation that appears in the traces recorded with both techniques is that upon creating the contact in most cases the conductance of the first contacts is higher than the conductance of the last contact before breaking in the preceding trace. As a final remark, we have checked that these materials do not form long atomic chains when stretching.

Our results agree with those of Berg {\it et al.}\cite{Berg14ElektrTransNanokontakteSeltenErdMetallen} who reported values of $(0.60\pm0.23)\,\frac{2e^2}{h}$ and $(0.83\pm0.32)\,\frac{2e^2}{h}$ for the last plateau right before breaking ($746$ curves) and first one after creating ($568$ curves) the contact, respectively, of Gd notched-wire MCBJs. Similarly, low conductance values were also observed for Dy ($(0.87\pm0.27)\,\frac{2e^2}{h}$ from $528$ breaking contact traces). Reported measurements on nanocontacts made out of Dy\cite{Muller10SwitchingConducDyNanocontactsMagnetostriction} showed non-trivial change of conductance as a function of the value of the external magnetic field. We observe similar but weaker magnetostrictive effect for some lithographic Gd MCBJ samples, in our case changing from sample to sample (see figure S1 at supplemental material).

From every set of traces with electrodes indentations that have not reached conductance values above $\approx20\,\frac{2e^2}{h}$, we build up histograms. Some of them are shown in Fig. \ref{fig:combined_hist_Gd_open_close_160805_01}. For most well-studied metals (Au, Pt, Ni\ldots) the position of the lowest maximum in the conductance histograms is very reproducible from experiment to experiment, although slight differences in the relative height and shape of the peaks have been reported\cite{Agrait93ConductanceStepsQuantizationAtomicSizeContacts,Smit01CommonOriginReconstructionChains,Calvo09KondoEffectFerromagneticAtomicContacts}.
In the case of Gd the position of such histogram peaks shows a lower reproducibility, which we attribute to different configurations of the electrodes. At Fig. \ref{fig:combined_hist_Gd_open_close_160805_01} we show five independent histograms (solid line, full color and vertical stripped lines patterns facilitate the identification of each one of them). We want to remark the strong resemblance between red (STM) and purple (MCBJ) breaking contact curves. On the other hand, for MCBJ closing contact higher conductance value is reached meaning that probably first MCBJ contacts are thicker. Yellow curve shows a very wide histogram, its corresponding traces were obtained with deeper indentations, meaning that a faster randomly rearrangement of the electrode tips were achieved. Black curve shows higher conductance values for both breaking and creating contact cases. The contacts that account for this result were poorly sharpened, meaning therefore that thicker tips are considered.

In addition, the last value of conductance before breaking contacts is well below that observed for other magnetic metals such as, e.g.,  Ni\cite{Calvo09KondoEffectFerromagneticAtomicContacts} where a mean value of conductance of $\approx1.5\,\frac{2e^2}{h}$ is found. With this background, it is remarkable that most of the lowest conductance peaks in the recorded histograms for Gd are well below $\frac{2e^2}{h}$.  

In order to gain further insight on the evolution of conductance traces, we have constructed 
intensity maps as a function of both the conductance and the displacement of the electrodes for our measured data, shown in Fig. \ref{fig:overlapped_traces_Gd}. 
This time, we collect all the conductance traces that we have measured with STM with indentations mostly up to $20\,\frac{2e^2}{h}$ and plot them in a two-dimensional histogram. In order to highlight the atomic-size contact area, we centre the traces at the same value of piezo displacement for a chosen conductance value (see figure caption). This way of representing data permits to check the dispersion of data at the low conductance stages unveiling the dependence of the most probable conductance on the applied strain. At the right-sided edge of the represented cloud of data a dispersion of $\approx0.25\,\frac{2e^2}{h}$ is clearly apparent, one quarter less than in the case of the conductance histograms in Fig. \ref{fig:combined_hist_Gd_open_close_160805_01}. Besides, from the dark-red branches and spots at the figure we can make the following interpretation: a monomeric atomic configuration (see calculations below) appears at both breaking and closing contacts, corresponding to the observed histogram value of $(0.9\pm0.3)\,\frac{2e^2}{h}$. However, a dimeric configuration is only present at breaking contact histograms, with a conductance of $(0.65\pm0.20)\,\frac{2e^2}{h}$.


\section{Theory}

Density functional theory (DFT) calculations are initially carried out with the LAPW code ELK. Correlations in the $f$ orbitals are treated using the DFT+U method in the Yukawa scheme\cite{Bultmark09MultipoleDecompoLDAUEnergyActinideCompounds} in the fully localized limit.  Spin-orbit coupling is treated in the non-collinear formalism. Within this framework, the bulk lattice constant matches the experimental one within $3\%$ deviation. To gain insight into the electronic structure in the constriction, we calculate the electronic structure of one-dimensional Gd chains and compare the differences driven by the reduced dimensionality. Chain structures are optimized in the lattice parameter. 

\begin{figure}[!htb]
\begin{center}$
\begin{array}{c}
\includegraphics[width=.5\textwidth]{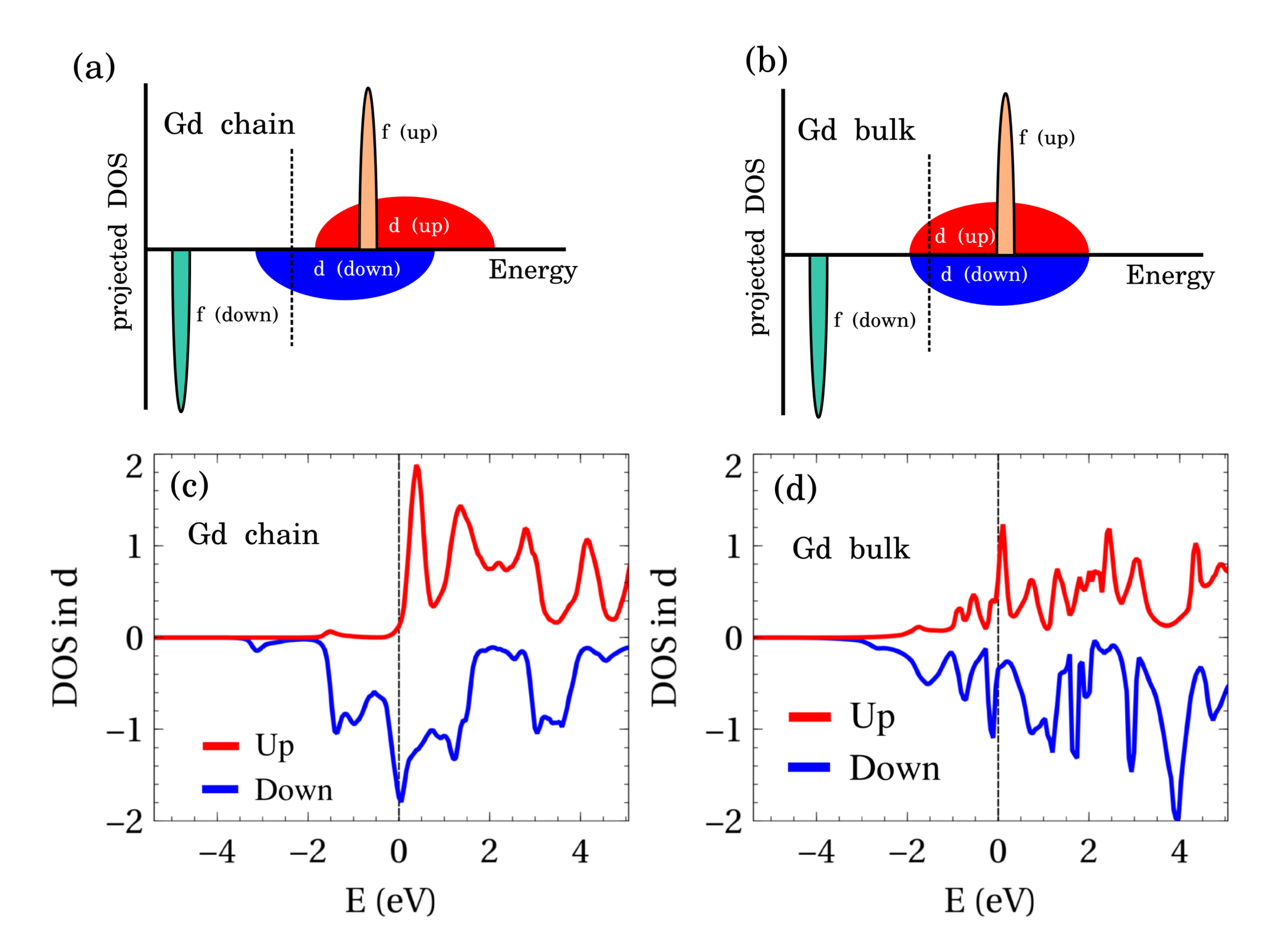} 
\end{array}$
\end{center}
\caption{Sketch of the projected DOS onto the $d$ and $f$ manifold for Gd in chain (a) and bulk (b). The $d$ manifold acquires a large spin splitting only in the chain case, while the $f$ manifold is spin polarized in both cases. Panels (c,d) show the DOS projected onto the $d$ manifold as calculated by first principles full potential method, for the chain (c) and bulk (d) Gd, in agreement with the sketch (a,b).}
\label{fig:dos_d_Gd}
\end{figure}

The magnetic structure strongly changes upon reducing the dimensionality from bulk to a one-dimensional chain (Fig. \ref{fig:dos_d_Gd}). Whereas the bulk structures yield a total magnetic moment of $\mu^{\rm bulk}=7.61\mu_B$, one-dimensional chains show larger moments of $\mu^{\rm chain}=8.9\mu_B$. 
The projected density of states reveals that, whereas bulk structures have spin polarization mainly coming from the $f$ levels plus a small contribution from $s$ ones (not shown), chains show a stronger polarization in the $d$ manifold arising from a stronger Stoner instability (see Fig. 5). This translates into energy differences between antiferromagnetic and ferromagnetic configurations, $J = E_{AF} -E_{FE}$, in chain structures which are much larger than the ones expected for bulk. Specifically we get $J= 0.5\,\mathrm{eV}$, favouring a fairly stable ferromagnetic configuration. Notice that the strong exchange coupling in chains can be understood as a consequence of the direct $d-d$ exchange coupling, whereas in bulk $d$ magnetism barely appears in our calculations.


Transport calculations for Gd ($<$111$>$) atomic contacts are also carried out. We choose a pyramidal form for both sides of the nanocontact, as expected in this type of experiments (see inset at Fig.  \ref{fig:3_traces_Gd}). The electron reservoirs, which make the nanocontact an open quantum system, are chosen to be Au electrodes. Au electrodes reduce the computational cost and introduce no artifact in the actual conductance of the Gd model nanocontacts when these contain a large enough number of Gd atoms.  The results presented below correspond to the minimum number of Gd atoms that needs to be considered (contacts with a larger number of Gd atoms have been studied, not finding significant differences). The transport methodology is the well-known DFT-based non-equilibrium Green's function formalism as implemented in the package ANT.G\cite{palacios2001fullerene,palacios2002first,louis2003implementing,jacob2011critical,jacob2006orbital}. This software uses the DFT functionality of Gaussian09\cite{frisch2009gaussian} to construct the one-particle Hamiltonian of the system. This Hamiltonian constitutes the basis for the implementation of the NEGF method through the Landauer-Keldysh formalism, which allows the simulation of open quantum systems connected to electron reservoirs.

\begin{figure*}[!htb]
  \centering
  \begin{tabular}{@{}p{0.225\textwidth}@{\quad}p{0.225\textwidth}@{\quad}p{0.225\textwidth}@{\quad}p{0.225\textwidth}@{}}
    \subfigimg[width=\linewidth]{(a)}{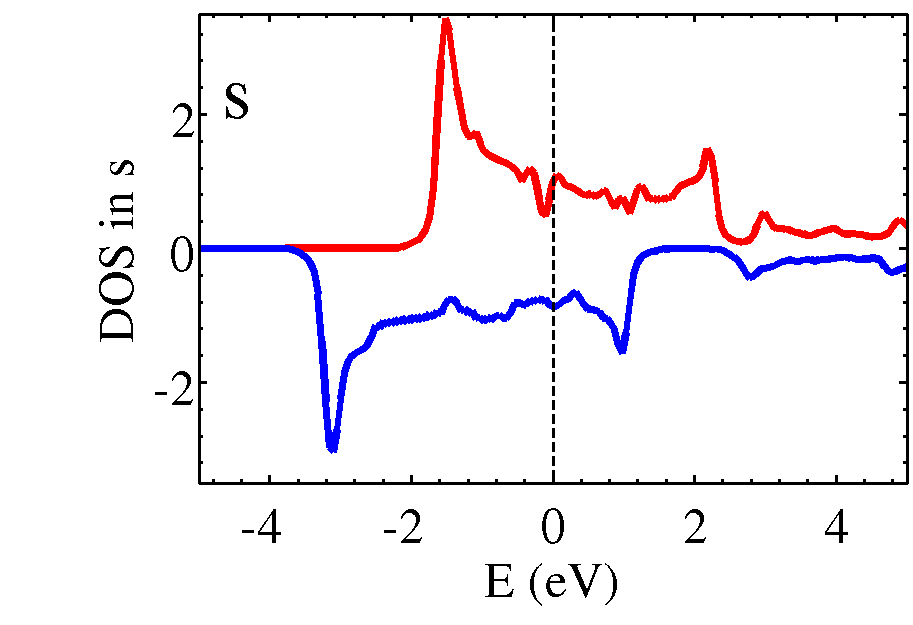} &
    \subfigimg[width=\linewidth]{(b)}{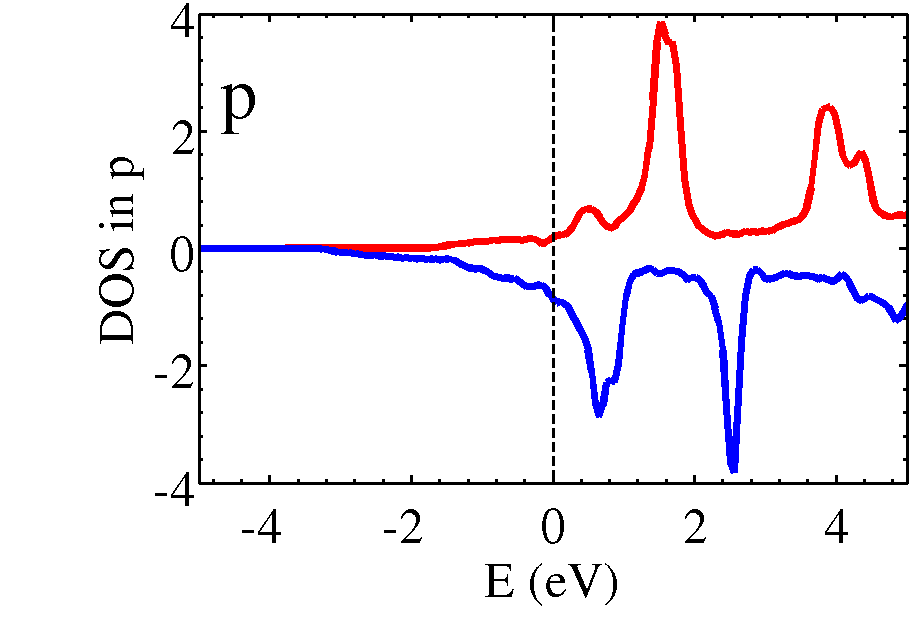} &
    \subfigimg[width=\linewidth]{(c)}{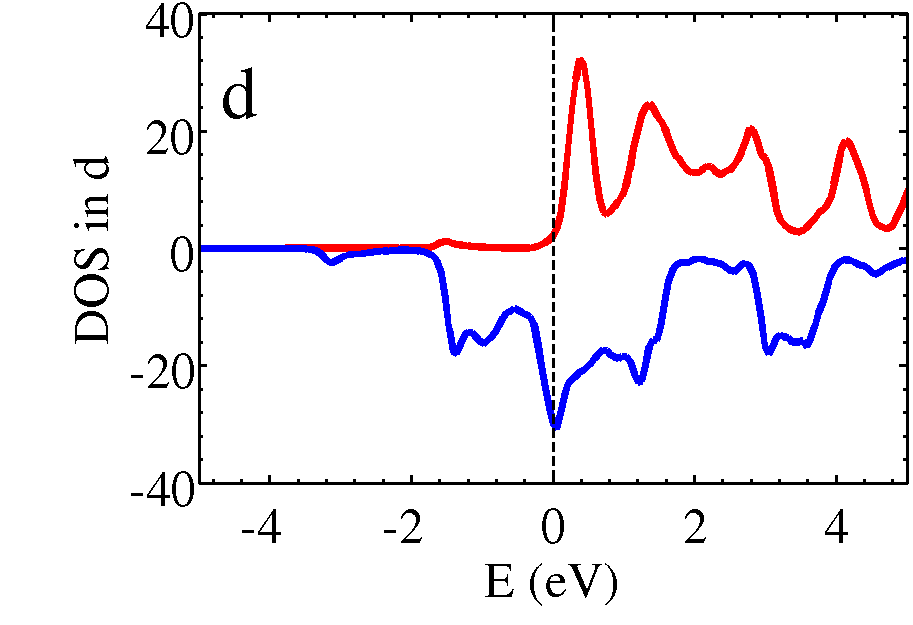} &
    \subfigimg[width=\linewidth]{(d)}{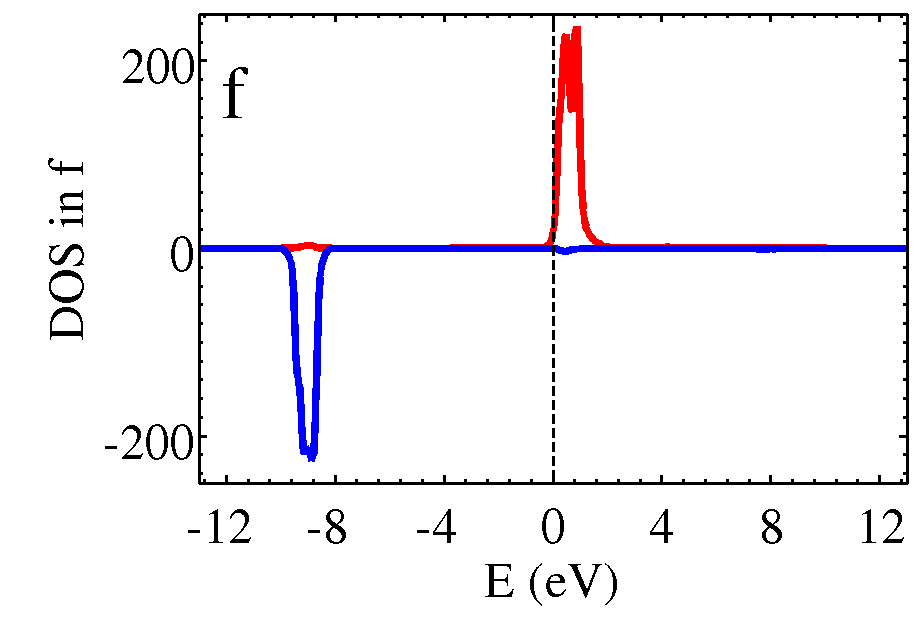} \\
    \subfigimg[width=\linewidth]{(e)}{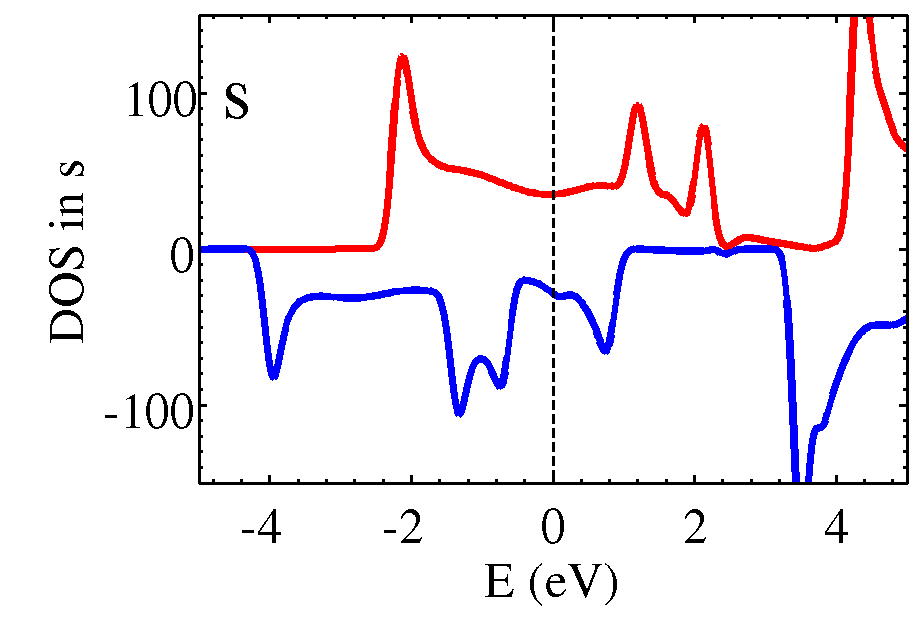} &
    \subfigimg[width=\linewidth]{(f)}{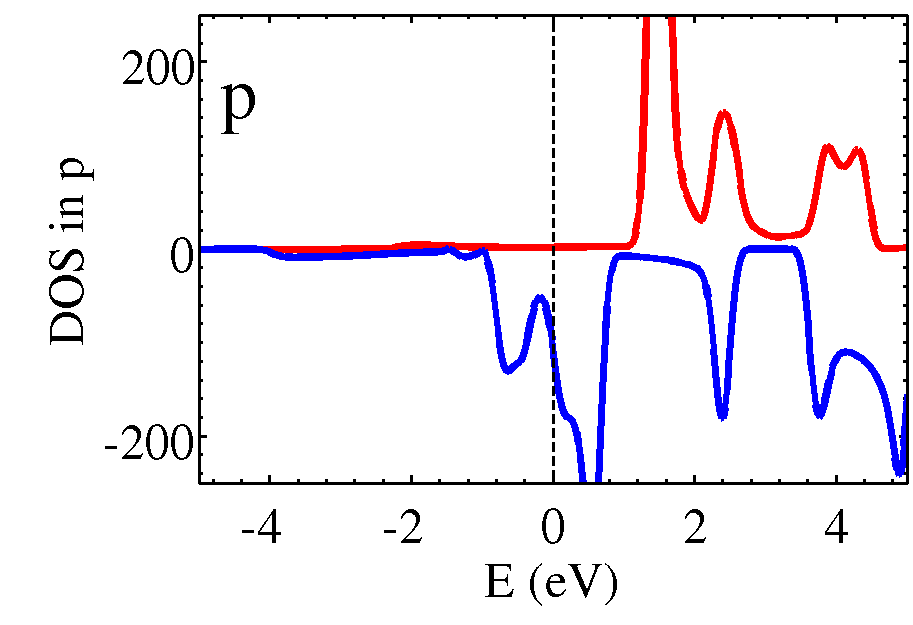} &
    \subfigimg[width=\linewidth]{(g)}{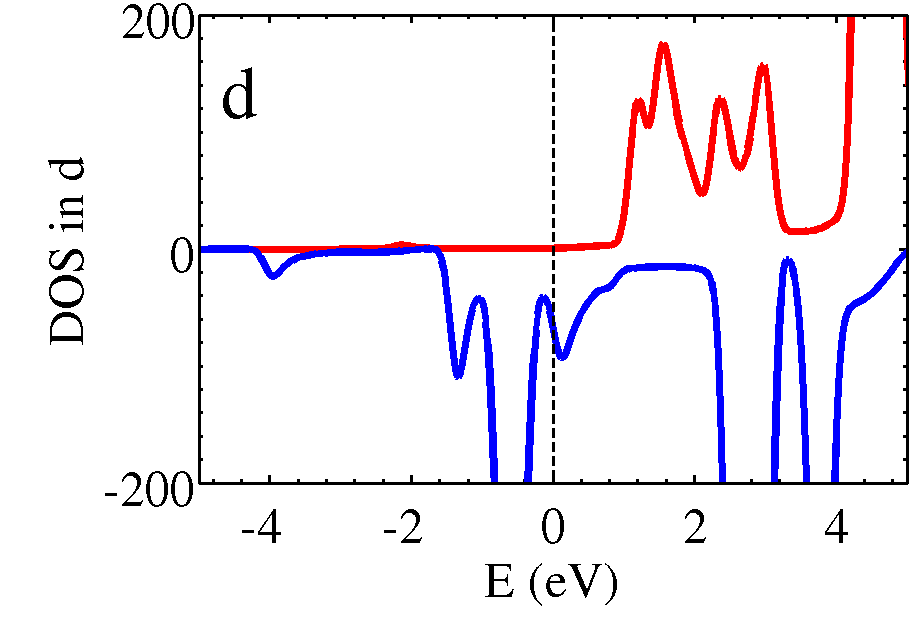} &
    \subfigimg[width=\linewidth]{(h)}{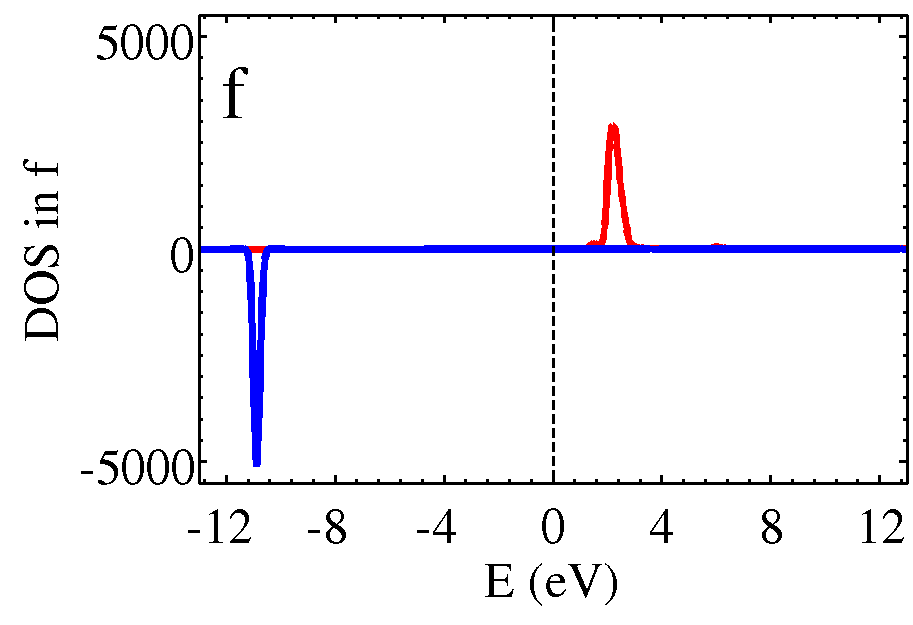}
  \end{tabular}
  \caption{Spin resolved density of states of Gd in chain structures calculated with ELK (LAPW) projected onto the manifolds $s$ (a), $p$ (b), $d$ (c) and $f$ (d). Same cases calculated with CRYSTAL14 (LCAO) ($s$ (e), $p$ (f), $d$ (g) and $f$ (h)).}
    \label{fig:dos_compared_Gd}
\end{figure*}



The NEGF-DFT method implemented in ANT.G operates in the framework of linear combination of atomic orbitals (LCAO). Therefore, as a necessary first check, we need to compare with the well-grounded results of ELK.
The basis ``Stuttgart RSC 1997 ECP'' has been used for Gd\cite{bergner1993ab,kaupp1991pseudopotential,dolg1993relativistic} in the LCAO-DFT calculations for chain and nanocontact structures. This basis set includes ECP (energy-consistent pseudopotentials) to describe the interaction with the core-electrons. Also, instead of using DFT+U to account for correlations in the $f$-orbitals, we make use of the hybrid functional HSE06\cite{heyd2004efficient}. In this hybrid functional developed for metals the exchange energy term is split into short-range  and long-range components and the Hartree-Fock long range is neglected but compensated by the PBE long range.

The suitability of our LCAO framework is confirmed, for instance, by comparing the DOS obtained for one-dimensional chains  with the ones obtained with ELK (see Fig. \ref{fig:dos_compared_Gd}). The LCAO calculations for infinite perfect chains are actually performed with the code CRYSTAL14\cite{dovesi2014crystal14} using the same basis set and functional as for the transport calculations. There we can see that, apart from a slight increase in the $s$ band widths, the DOS for all $s,p$, and $d$ manifolds share most important features in the energy window shown. The splitting of the $f$ bands is of the same order in both approaches as well ($\approx 9\,\mathrm{eV}$) and the positioning of the peaks with respect to the Fermi level is similar. The LCAO method reveals a total magnetic moment similar to the value obtained with LAPW-ELK.


\begin{figure}[!htb]
\begin{center}$
\begin{array}{c}
\includegraphics[width=0.99\linewidth]{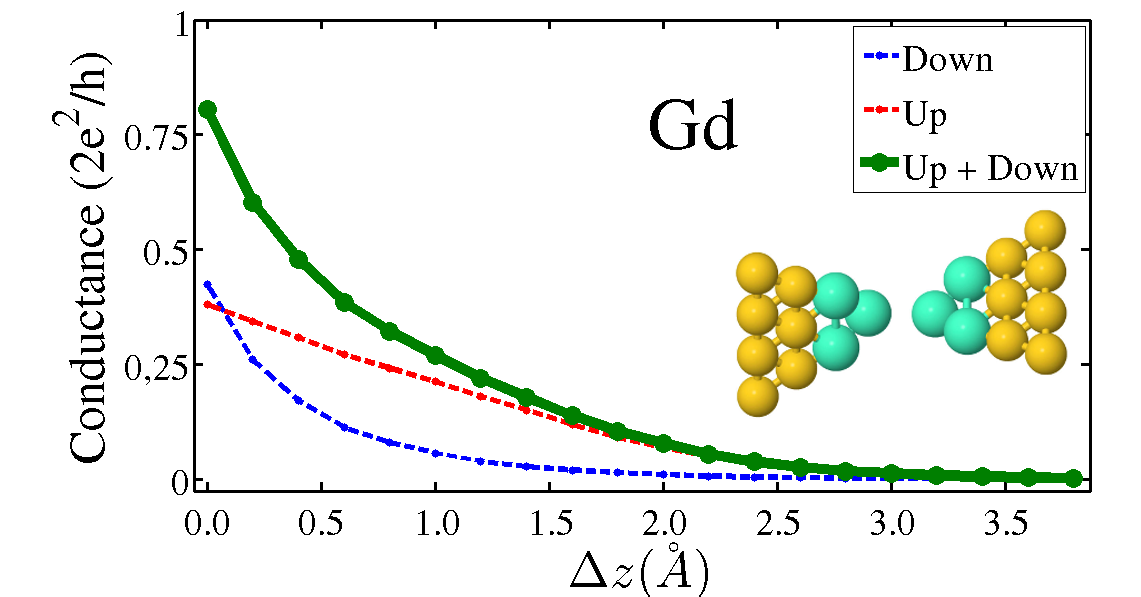}\\
\includegraphics[width=0.99\linewidth]{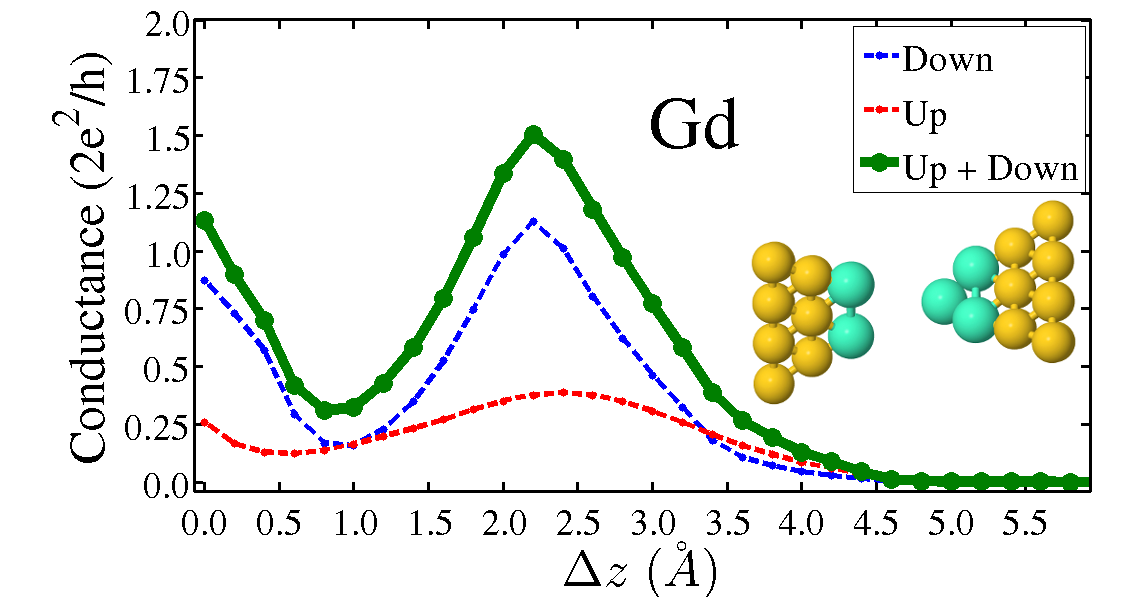}
\end{array}$
\end{center}
\caption{Gd $<$111$>$ nanocontact calculated conductance-distance characteristic. Upper plot: two atomically-sharp tips in the $<$111$>$ direction forming a dimeric contact. The distance $\Delta z$ equals $0$ when the distance between both tip apex atoms equals the n.n. distance in bulk. Lower plot: an atomically-sharp tip  in the $<$111$>$ direction against a blunt $<$111$>$ one. Both tips form a monomeric contact. The distance $\Delta z$ equals $0$ when the distance between the tip apex atom in the atomically-sharp tip and each tip apex atom in the blunt one equals the n.n. distance in bulk.}
\label{fig:characteristic_Gd}
\end{figure}



The calculated conductance-distance characteristics for Gd nanocontacts are shown in Fig. \ref{fig:characteristic_Gd} for both monomer and dimer configurations (see insets in Fig. \ref{fig:characteristic_Gd}). As anticipated in the discussion of the experimental results, monomer and dimer configurations are expected to form when breaking the contacts while in most of the cases only monomers are expected to appear when forming the contacts.  The piezo displacement is simulated by opposite displacement of the two tips, while keeping their atomic structure intact. Thus, we do not make any distinction between breaking and creating contacts, although a small difference in the average atomic bond distance is expected between the two processes if relaxation were allowed. Relaxation would also permit to simulate the plasticity effects (jumps in conductance), as seen in this type of experiments. This is, however, computationally too costly and beyond the scope of our discussion here.

As their periodic counterparts (bulk and chains), Gd nanocontacts show a purely ferromagnetic behaviour all along the breaking process. Anti-parallel magnetic configurations (between the two tips) show smaller conductance values, but these magnetic states have a higher energy and tend to relax into the ferromagnetic ones. The current is spin-polarized with a dominant contribution from the minority channel (spin-up here in red) for stretched dimeric contacts and from the majority one (spin-down in blue) for monomeric contacts. The calculated total conductance at bulk near-neighbour distance between tip apex atoms (or zero displacement, $\Delta z=0$) is $0.80\,\frac{2e^2}{h}$ for the dimer and $1.15\,\frac{2e^2}{h}$ for the monomer contacts. Both values, along with the ones nearby for small displacements (representing actual stretched contacts) fall within their tentatively assigned experimental bright spots seen in Fig. \ref{fig:characteristic_Gd}. Notice that, due to lack of relaxation in our calculations, longer displacements may not represent actual atomic configurations since sudden plastic deformations must occur. Remarkably, even so, for the monomer configuration we obtain an increase of conductance as the tip-tip distance increases (see Fig. \ref{fig:characteristic_Gd}(b)), as experimentally observed sometimes.

\begin{figure}[!htb]
  \centering
\begin{tabular}{@{}p{0.475\textwidth}@{}}
    \subfigimg[width=\linewidth]{(a)}{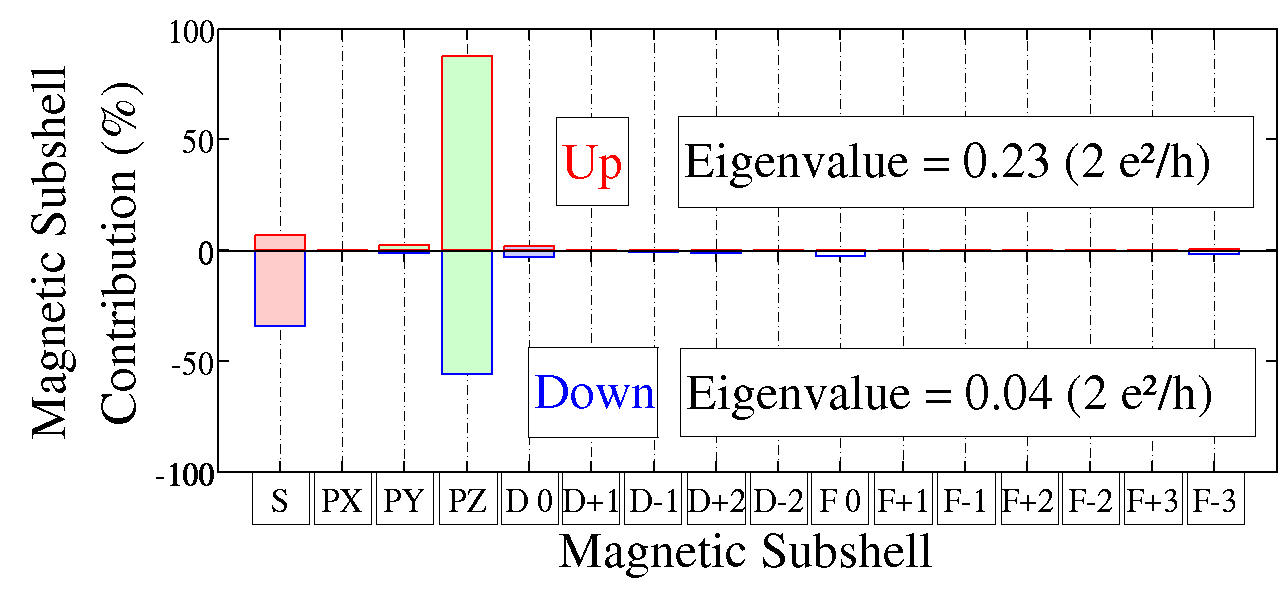} \\
    \subfigimg[width=\linewidth]{(b)}{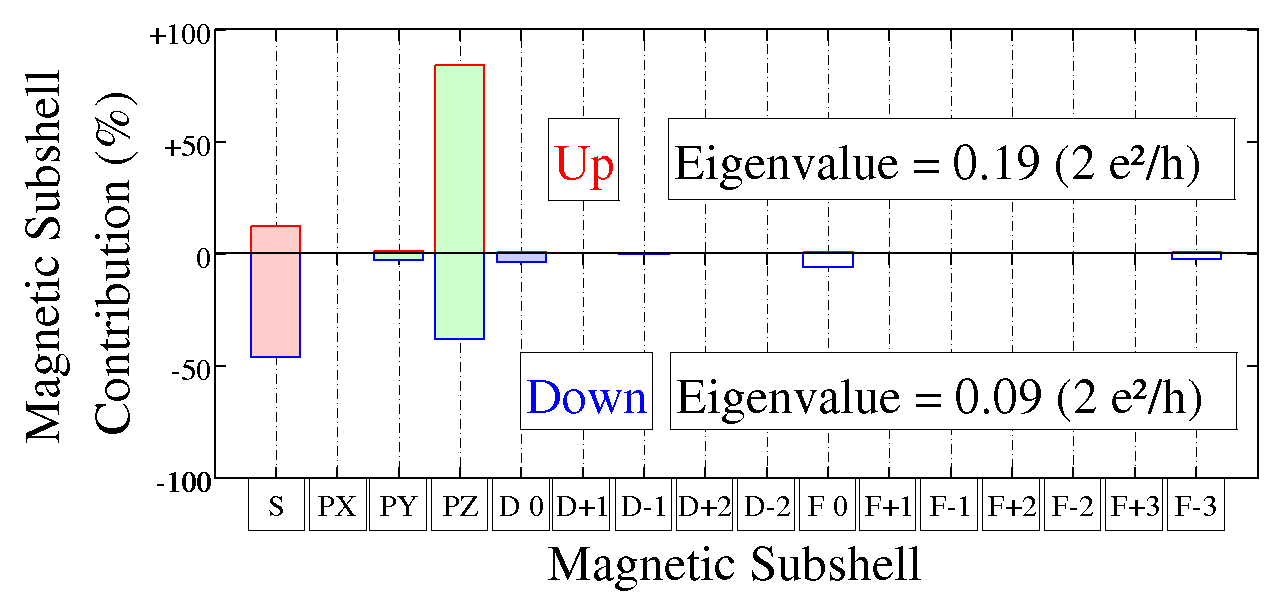} \\
    \subfigimg[width=\linewidth]{(c)}{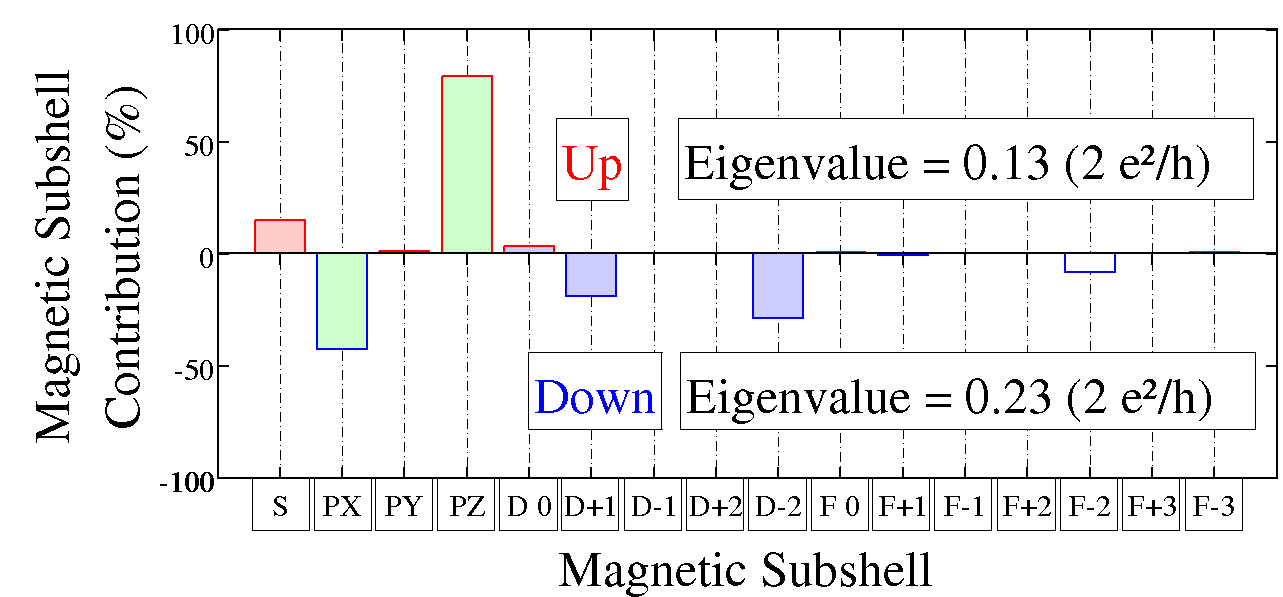}
  \end{tabular}
\caption{(a) Spin resolved principal eigenchannel projected onto all the magnetic shells of a Gd dimeric nanocontact at a tip-tip displacement of $1.0\,\mathrm{\AA}$; (b) and (c) the same but for a  monomeric contact at tip-tip displacement of $\Delta z = 1.0\,\mathrm{\AA}$ and $\Delta z = 2.2\,\mathrm{\AA}$, respectively. ``Up'' electrons correspond to the ``minority'' spin component while ``down'' electrons to the ``majority'' one.}
\label{fig:channelsGd}
\end{figure}

A deeper insight into the  electronic nature of transport can be revealed by analyzing the nature of the eigenchannels\cite{jacob2006orbital} involved in the conductance. In Fig. \ref{fig:channelsGd} we plot the conductance of the spin up and spin down dominant eigenchannels  for three representative examples. We have chosen: (a) a dimer contact at a displacement of  $\Delta= 1.0\,\mathrm{\AA}$ and (b) a monomer contact at a displacement of $\Delta= 1.0\,\mathrm{\AA}$ and (c) $\Delta= 2.2\,\mathrm{\AA}$. In general, the eigenchannels do not show a dominant $s$ character. For the case (a) they display mostly a $p_z$ character (minority) or $sp_z$ character (majority). As in the case of Al nanocontacts\citep{Cuevas1998EvolConducChanMetalAtContacElastDeform,Scheer97ConducChannTransAtomSizeAlContacts}, this $sp_z$ hybridization seems to account for the low conductance in contrast to the one expected for pure $s$ channels. In the monomer case, in addition to the $sp_z$ hybridization\citep{Cuevas1998EvolConducChanMetalAtContacElastDeform,Scheer97ConducChannTransAtomSizeAlContacts}, the transition from $sp_z$-like eigenchannels at smaller displacements (Fig. \ref{fig:channelsGd}(b)) to an eigenchannel with a strong $d$ character for majority spins (Fig. \ref{fig:channelsGd}(c)) at larger displacements seems to also play a role in the change in the slope of the conductance-displacement curve, as seen in Fig. \ref{fig:characteristic_Gd}(b). Finally, we should mention that, at least at zero bias, no direct contribution from $f$ orbitals seems to appear in transport, apart from enhancing the spin polarization in the other subshells.


\section{Conclusions}

We have carried out electrical current measurements on atomic-size contacts made out of Gd under cryogenic conditions. Unlike the case of $3d$  ferromagnetic materials and despite the $d^1$ valence of Gd, their single atom conductance is typically smaller than $\frac{2e^2}{h}$. This is might be at first sight surprising, because  in both cases there is a wide $s$-band at the Fermi energy, which normally provides a transmissive channel and which, along with an additional contribution coming from the $d$ channels,  may give conductance values above $\frac{2e^2}{h}$. The results are reproducible for both STM and MCBJ measurements in all details: lengths of plateaus, conductance shapes of plateaus\ldots.

In agreement with the experiments, our DFT calculations generically give conductance values smaller than $\frac{2e^2}{h}$, with differences between monomer and dimer configurations.  The analysis of the eigenchannels shows that this is due to a hybridisation of the $s$ and $p_z$ channels, which reduces the conductance of a pure $s$ channel. This is also in line with the increasing conductance on the last plateau as the electrodes are pulled apart, which has also been observed in Al atomic contacts and for which the $sp_z$ hybridisation is also known to play a role. In this case, the $d$ orbitals also seem to play a role in this conductance rise. Finally, our zero-bias measurements do not seem to be strongly influenced by the large $f$ local magnetic moments, firstly because $f$ electrons do not participate in conduction and, secondly, the stable ferromagnetic configuration will avoid electric conductance variations due to magnetic disorder.


\begin{acknowledgments}

BO, CS, JFR, JJP and CU acknowledges  financial support by MEC-Spain  (FIS2013-47328-C2) and  the Generalitat Valenciana under grant no. PROMETEO/2012/011. CS and JJP acknowledge the EU structural funds and the Comunidad de Madrid MAD2D-CM program under grant nos. S2013/MIT-3007 and S2013/MIT-2850. CS and JJP also acknowledge the computer resources and assistance provided by the Centro de Computaci\'on Cient\'ifica of the Universidad Aut\'onoma de Madrid and the RES. JLL and JFR acknowledge Marie Curie ITN  SPINOGRAPH FP7 under REA grant agreement  607904-13. BO acknowledges financial support by MEC-Spain (FIS2010-21883-C02-01) under briefs stays abroad scholarship.

We gratefully acknowledge helpful discussion with C. S\"urgers and M. Fonin regarding the material properties of rare earth metals and for providing the Gd material. We thank to C. Sabater, F. Strigl and H. Ballot for experimental help.

\end{acknowledgments}

\appendix

\bibliography{berbiblio}

\end{document}


\newcommand{\bra}[1]{\left\langle#1\right|}
\newcommand{\ket}[1]{\left|#1\right\rangle}
\newcommand{\bracket}[2]{\big\langle#1 \bigm| #2\big\rangle}
\newcommand{\PPV}{\ket{0_{\rm PP}}}
\newcommand{\Tr}{{\rm Tr}}
\renewcommand{\Im}{{\rm Im}}
\renewcommand{\Re}{{\rm Re}}
\newcommand{\MC}[1]{\mathcal{#1}}
\newcommand{\p}{{\prime}}
\newcommand{\pp}{{\prime\prime}}
\newcommand{\ppp}{{\prime\prime\prime}}
\newcommand{\pppp}{{\prime\prime\prime\prime}}
\newcommand{\TK}{T_{\rm K}}

\title{Electronic Transport in Gadolinium Atomic-Size Contacts\\(SUPPLEMENTAL  MATERIAL)}

\author{B. Olivera}
\affiliation{Departamento de F\'isica Aplicada, Facultad de Ciencias, Universidad de Alicante, San Vicente del Raspeig, E-03690, Spain.}
\author{C. Salgado}
\affiliation{Departamento de F\'isica de la Materia Condensada, Universidad Aut\'onoma de Madrid, E-28049 Madrid, Spain.}
\author{J. L. Lado}
\affiliation{International Iberian Nanotechnology Laboratory (INL), 4715-330 Braga, Portugal.}
\author{A. Karimi}
\affiliation{Department of Physics, University of Konstanz, Universit\"{a}tsstra{\ss}e 10, D-78464 Konstanz, Germany.}
\author{V. Henkel}
\affiliation{Department of Physics, University of Konstanz, Universit\"{a}tsstra{\ss}e 10, D-78464 Konstanz, Germany.}
\author{E. Scheer}
\affiliation{Department of Physics, University of Konstanz, Universit\"{a}tsstra{\ss}e 10, D-78464 Konstanz, Germany.}
\author{J. Fern\'andez-Rossier}
\affiliation{Departamento de F\'isica Aplicada, Facultad de Ciencias, Universidad de Alicante, San Vicente del Raspeig, E-03690, Spain.}
\affiliation{International Iberian Nanotechnology Laboratory (INL), 4715-330 Braga, Portugal.}
\author{J. J. Palacios}
\affiliation{Departamento de F\'isica de la Materia Condensada, Condensed Matter Physics Center (IFIMAC), and
Instituto Nicol\'as Cabrera, Universidad Aut\'onoma de Madrid, E-28049 Madrid, Spain.}
\author{C. Untiedt}
\affiliation{Departamento de F\'isica Aplicada, Facultad de Ciencias, Universidad de Alicante, San Vicente del Raspeig, E-03690, Spain.}

\date{\today}

\pacs{73.63.Rt, 71.15.Mb, 72.25.Ba}

%
%

\maketitle

\section{Magnetostriction in Gd nanocontacts}

\begin{figure*}[!htb]
\begin{center}$
\begin{array}{cc}
\includegraphics[width=3.in]{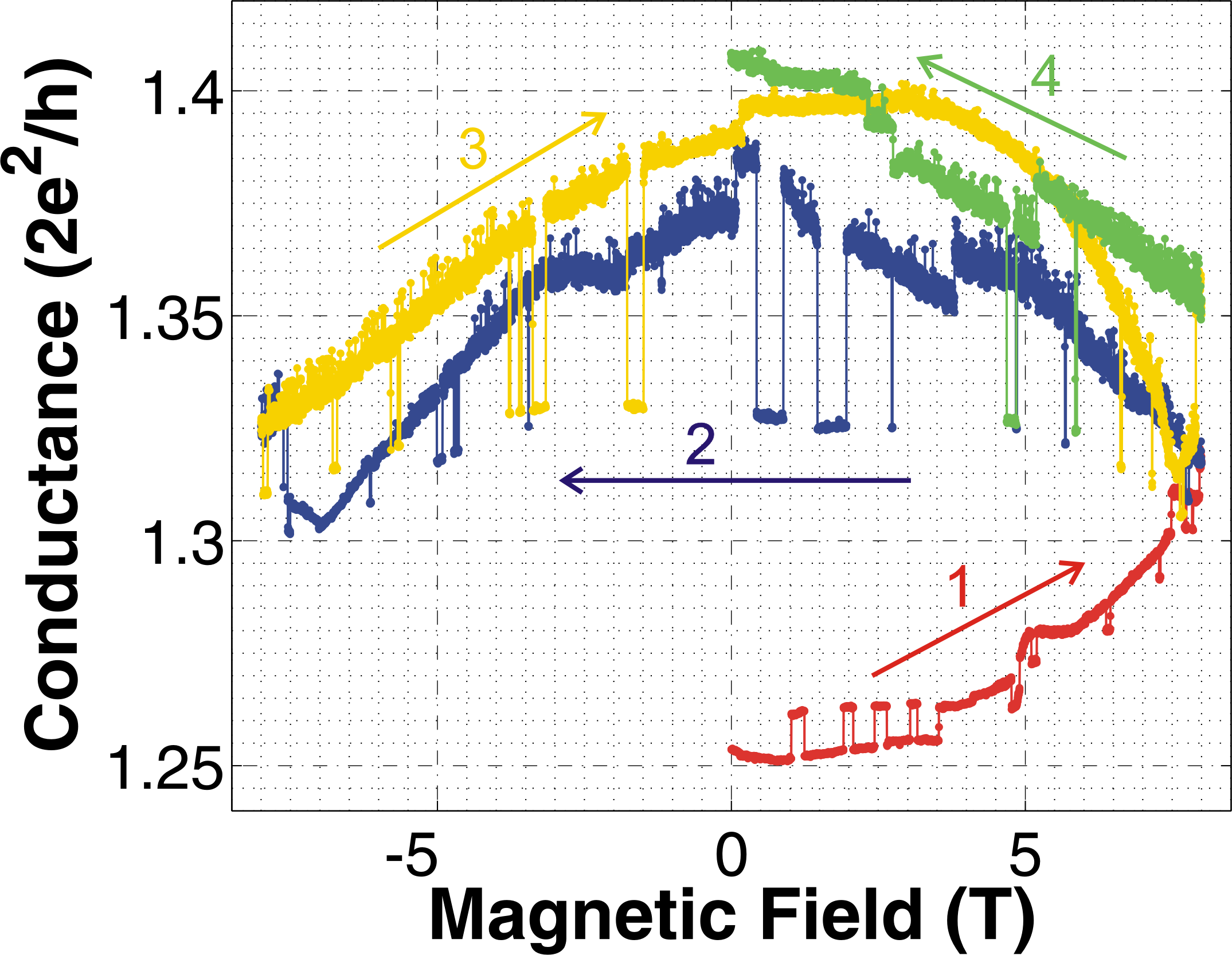} &
\includegraphics[width=3.in]{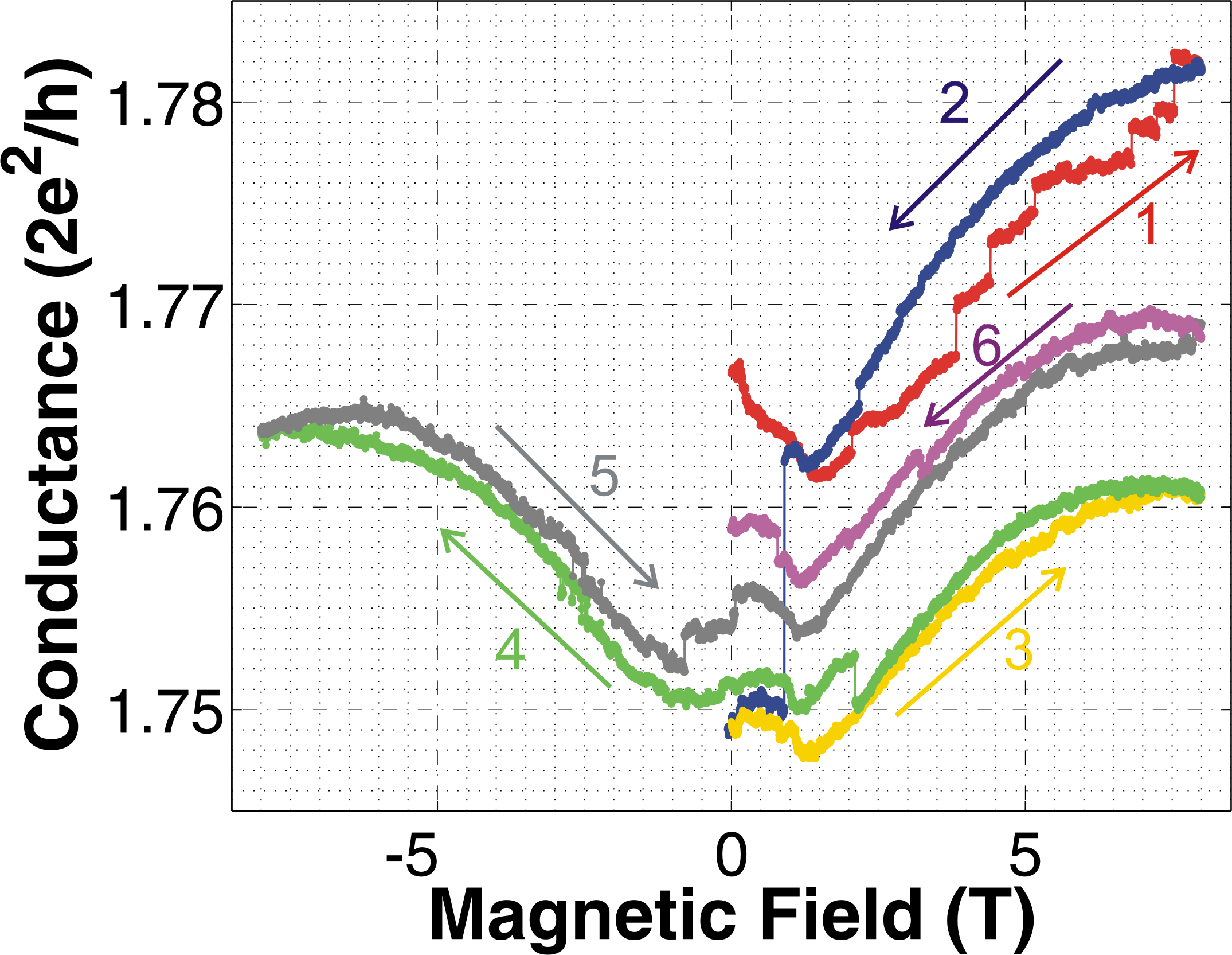} \\
\includegraphics[width=3.in]{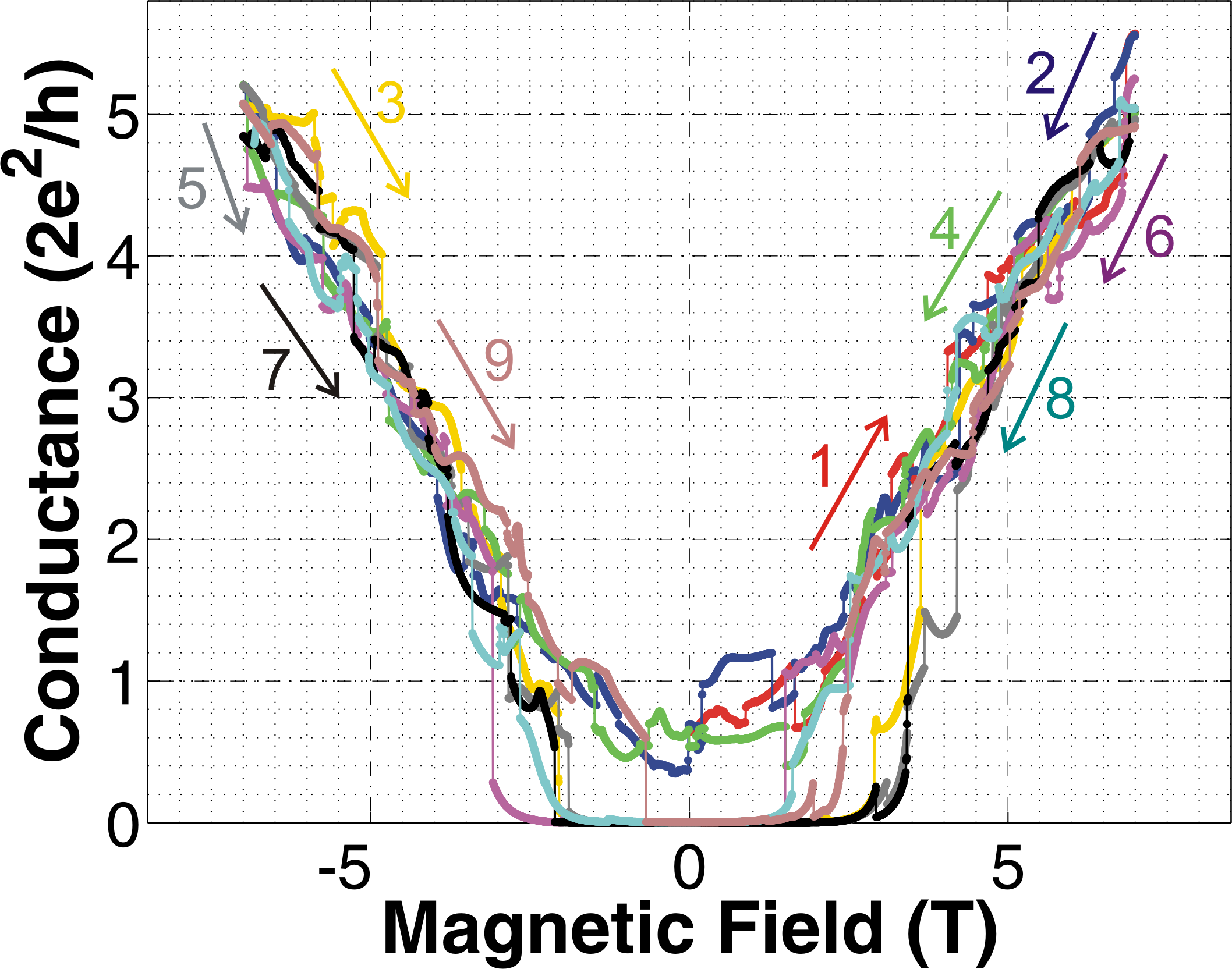} &
\includegraphics[width=3.in]{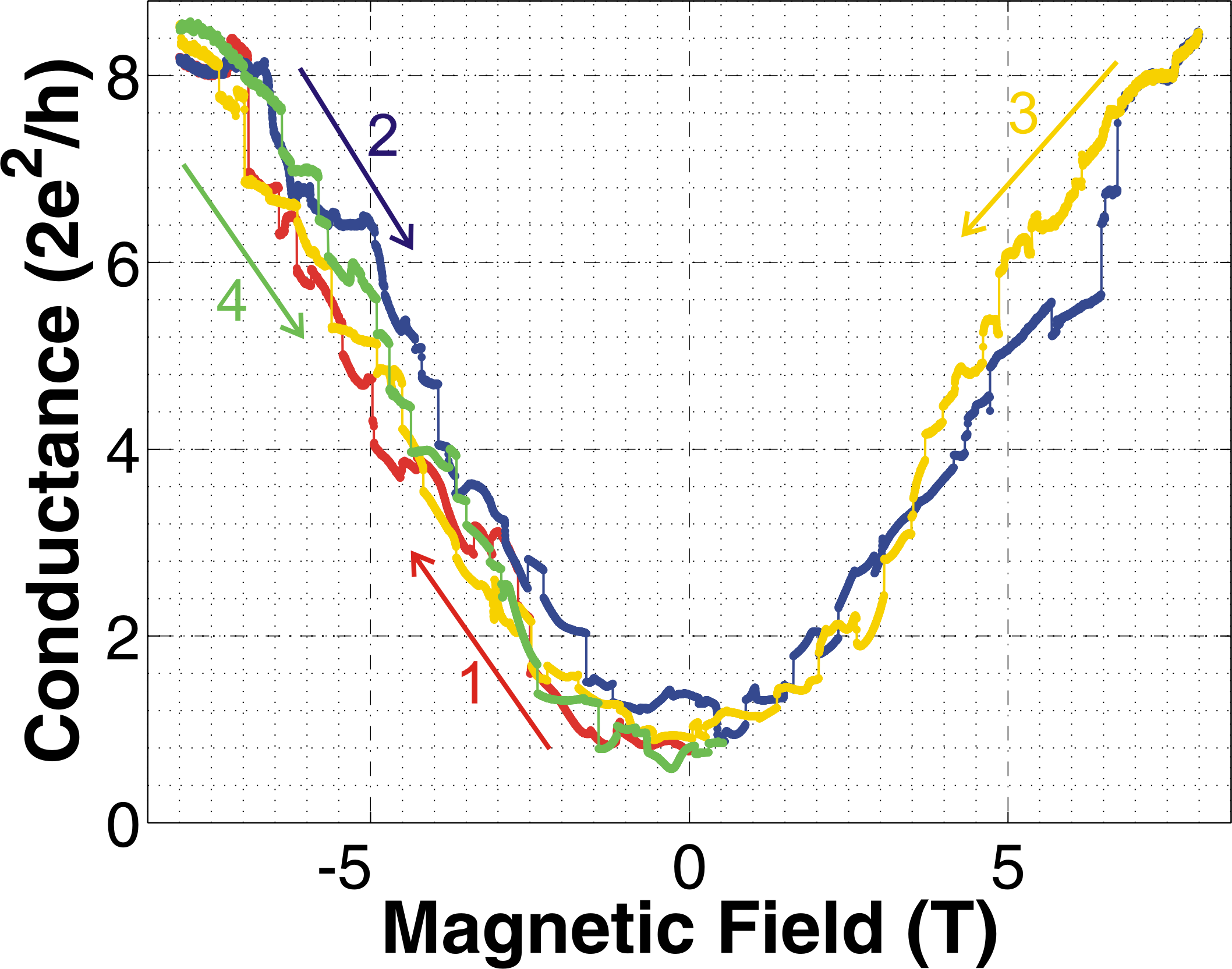} \\
\end{array}$
\end{center}
\caption{Magneto-conductance measurements for atomic-size contacts made out of Gd (see text). Measurements have been taken with MCBJ under cryogenic conditions (vacuum below $10^{-5}\,\mathrm{mbar}$ and temperature in equilibrium with liquid He). Both plots at the upper part of this figure correspond to different contacts of the same sample, whereas the other two at the lower part correspond to another one. Coloured  arrows and numbers indicate the sweeping order for every run. Magnetic field direction is always the same, that is pointing out of plane of the substrate (the sign indicates the sense of the field). Sweeping speed is $0.2\,\mathrm{T/min}$, except for cases shown at upper-left plot, where speed is $0.1\,\mathrm{T/min}$.}
\label{fig:magnetoconductance}
\end{figure*}

We have performed magneto-conductance measurements for different samples. At Fig. \ref{fig:magnetoconductance} we show our main results. We find that the two samples (see figure caption) show very different magnetostrictive behaviour. The pristine Gd pellets for each sample come from different batches, respectively. Moreover, evaporation parameters are slightly different for each case.
%
%
%
We see that for one sample (upper plots) conductance changes with a rate of $\approx0.01\,\frac{2e^2}{h \cdot T}$, even less. At variance, the magnetostriction of the other sample (lower plots) changes with a rate up to $\approx0.7\,\frac{2e^2}{h \cdot T}$, that is much higher than the other one. It is interesting to see how the magnetoconductance traces shown in the lower plots resemble conductance-distance traces shown in the main text. We thus interpret the observation as being caused by magnetostriction as observed earlier for Dy\cite{Muller10SwitchingConducDyNanocontactsMagnetostriction}. This means that the mechanical displacement of the electrodes is provoked by the external magnetic field and not by the pushing rod of the MCBJ. While Dy reveals a strong magnetostriction in all crystalline directions, in Gd the bulk magnetostriction is much smaller and in addition direction dependent with opposite sign in different directions\cite{Nigh63MagnetElecResistGdSingleCrystals,Coleman64VolMagnetostricGdSinglCrystals}. We argue that depending on the evaporation rate the crystallinity of the films varies resulting in a strong variation of the magnetostriction of the nanobridge. Further studies are necessary to verify this suggestion.

\appendix

\bibliography{berbiblio}